\documentclass[fleqn,usenatbib]{mnras}

\usepackage{newtxtext,newtxmath}

\usepackage[T1]{fontenc}

\DeclareRobustCommand{\VAN}[3]{#2}
\let\VANthebibliography\thebibliography
\def\thebibliography{\DeclareRobustCommand{\VAN}[3]{##3}\VANthebibliography}

\usepackage{graphicx}	
 
\usepackage{amsmath}	
\usepackage{amssymb}	
\usepackage{subcaption}
\usepackage{url}
\usepackage{color}

\title[Tracing ABH Seeds and PBHs with LISA-Taiji joint network]{Tracing Astrophysical Black Hole Seeds and Primordial Black Holes with LISA-Taiji Network}
\author[Y. Yang et al.]{
Yuchan Yang $^{1}$,\thanks{yangyuchan02@gmail.com}
Wen-Biao Han$^{2,1,3,5},$\thanks{conressponding author:  wbhan@shao.ac.cn}
Qianyun Yun $^{1}$,
Peng Xu$^{4,2}$,
and Ziren Luo$^{2,4}$
\\
$^{1}$Shanghai Astronomical Observatory, Chinese Academy of Sciences, Shanghai 200030, China\\
$^{2}$Hangzhou Institute for Advanced Study, University of Chinese Academy of Sciences, Hangzhou 310124, China\\
$^{3}$School of Astronomy and Space Science, University of Chinese Academy of Sciences, Beijing 100049, China\\
$^{4}$Institute of Mechanics, Chinese Academy of Sciences, Beijing 100190, China\\
$^{5}$ Shanghai Frontiers Science Center for  Gravitational Wave Detection, 800 Dongchuan Road, Shanghai 200240, China}

\date{Accepted XXX. Received YYY; in original form ZZZ}

\pubyear{2021}

\begin{document}
\label{firstpage}
\pagerange{\pageref{firstpage}--\pageref{lastpage}}
\maketitle

\begin{abstract}
In this work, we discuss the improvement that the joint network of LISA and Taiji could provide on exploring two kinds of black hole formation mechanisms. For astrophysical origin, we consider light seed and heavy seed scenarios, and generate populations accordingly. We find that the joint network has the potential to observe growing light seeds in the range $15< z < 20$ while single detector can hardly see, which would shed light on the light seeding mechanism. For the heavy seeds, the joint network only improves the signal-to-noise ratio. For primordial origin, we calculate the event rate at $z > 20$ and detection rates of LISA and the joint network. The joint network expands LISA’s horizon towards lower mass end, where the event rate is high, so we have better chance observing primordial black holes with the joint network. We also estimate the parameters using Fisher matrices of LISA and the joint network, and find that the joint network significantly improves the estimation.
\end{abstract}




\section{Introduction}

The undergoing ground-based gravitational wave (GW) detection projects such as LIGO, Virgo, and KAGRA have enabled us to ‘hear’ from compact objects in the deep sky through their inspiral, coalescence, and ringdown. In addition, collaborations between these GW detectors will provide us more information that single detector may hardly achieve, which has shed light on some open questions such as the black hole (BH) seed formation mechanism and the abundance of primordial black holes (PBHs) \citep{2020ARAAInayoshi,2021NatRPvolonteri}.

In the next decade, the future GW detectors would be able to tell us more. On the one hand, ground-based GW detectors, such as the Einstein Telescope \citep[ET;][]{ET} and Cosmic Explorer \citep[CE;][]{CE} will be able to detect the signal from coalescing stellar-origin BHs out to very large redshifts $z\sim 20$, and also will be able to discover signals from merging PBHs if they form \citep{ChenHuang2020JCAP}. On the other hand, space-borne GW detectors such as Laser Interferometer Space Antenna \citep[LISA;][]{2017arXiv170200786A}, Taiji \citep{hu2017taiji} and TianQin \citep{2016CQGra..33c5010L}, which are sensitive to lower frequencies around a milli Hz, will be able to detect signals from heavier, massive black hole binaries of $10^4 M_\odot -10^7 M_\odot$ out to comparable redshifts $z > 20$, and stellar BHs in the near Universe doing their secular inspiral and on their track to be detected by ground-based interferometers \citep{Sesana2016PhRvL.116w1102S}.

Collaborations between these detectors will provide us with better results such as significant improvement on sky localization and polarization determination \citep[e.g.][for LISA-TianQin network]{shuman2021massive}. On the LISA-Taiji network, \citet{2020NatAs...4..108R} and \citet{2020PhRvD.102b4089W} showed the sky localization improvement of the joint observation compared with LISA alone. \citet{omiya2020searching, seto2020gravitational, orlando2021measuring} measured the LISA-Taiji capabilities for the stochastic GW background observation. \citet{wang2020hubble, wang2021forecast} showed the impact on cosmological parameter determination with the joint observation. \citet{wang2021observing} valuated the observation constraints on the GW polarizations from the joint observation. \citet{2021PhRvD.104b4012W} discussed the performances of different LISA-Taiji configurations. Following these works, we consider the collaboration between LISA and Taiji, and discuss the improvements that this joint network could provide in the discovery of BH seed formation and exploration of PBHs before cosmic dawn.

The formation and evolution mechanism of BH seeds is not well known yet. Several scenarios have been proposed. In the ‘light seed’ scenario, metal-free/poor population III (popIII) stars collapse into BHs with masses of a few $10^2 M_{\odot}$ in the early Universe at $z \sim 20-30$ \citep[e.g.][]{madau2001massive}. In the ‘heavy seed’ scenario, seeds form from the direct collapse of massive gas clouds in dark matter haloes of typical masses of $10^8$ solar masses, illuminated by UV radiation to prevent cooling by molecular hydrogen and prevent fragmentation \citep{2020ARAAInayoshi}. Seeds of supermassive BHs (SMBHs) powering high-redshift quasars might have two origins: they might be of astrophysical origin [astrophysical black holes (ABHs)] resulting, e.g. from the collapse of very massive stars, or of primordial origin (PBHs) formed in the radiation-dominated era. Seeds are transient populations that formed at very high redshifts to account for the presence of million to billion solar-mass SMBHs shining as quasars. Once the seeds are formed, they grow either via gas accretion, or more importantly through coalescence with other seeds\citep[e.g.][]{volonteri2003assembly}. This implies that BH seeds may form binaries and become high-z GW sources. As \citet{2021MNRAS.500.4095V} pointed out, with the future detectors such as ET and LISA, we should be able to observe BH seeds during their formation and growth, and it would be ground breaking. However, there are still parts of growing light seeds that lie out of the horizon of LISA. A joint observation of LISA and Taiji would broaden the horizon, leading to more detection of growing light seeds, if they really exist. LISA and Taiji can detect the merger of binaries with mass as low as $10^3 m_\odot$ and redshift more than 20, which the current ground detectors like LIGO and the third-generation detectors may not achieve.


 The compact objects form in the radiation-dominated early uni- verse out of the collapse of large overdensities, and form binaries via random gravitational decoupling from the Hubble flow before the matter-radiation equality \citep[see][for a recent review]{sasaki2018primordial}. PBHs of different masses could contribute large fractions of dark matter halo, which motivates efforts in searching for PBHs. It is plausible that current BH population is a mixture of ABHs and PBHs. However, at $z$ > 30, all BHs are of primordial origin \citep{2021JCAP...05..003D}, since ABHs are formed only at lower $z$. LISA is capable of detecting GWs from $z$ > 30, thus should be able to detect PBHs at high $z$. In addition, a joint network of LISA and Taiji could not only increase the detection rate of PBHs, but also improve the accuracy on parameter measurement, which would be necessary in identifying the primordial origin.

This paper is organized as follows. In section \ref{sec.ABH}, we explain how we generate the BH seeds, the model we use, and how we calculate the signal-to-noise ratio. In section \ref{sec.PBH}, we explain how we generate the PBH population and how we simulate the GW signal. Then we calculate the event rate of PBH and detection rate of LISA and the joint network. Finally, we use Fisher information matrix to compare the parameter estimation ability of LISA and the joint network. In section \ref{sec.result}, we gather the results from last two sections, and discuss the improvement of the joint network.

\section{Astrophysical Black Holes}\label{sec.ABH}
One popular theory of BH formation is the so-called BH seeding mechanism. The BHs formed in this way are called ABHs. In this theory, BH seeds are formed from remnants of different astrophysical bodies. Then they grow via accretion of surrounding gas and coalescence with other seeds. The simulation work of ABH population and evolution can be found in \citet{2016PhRvD..93b4003K, 2021MNRAS.500.4095V}.

\subsection{ABH population}
Following \citet{2016PhRvD..93b4003K}, we employ ABH binary catalogues using the model of \citet{barausse2012evolution} (Catalogs are downloaded from https://people.sissa.it/$\sim$barausse/catalogs/). The model follows the evolution of baryonic structures along a dark matter merger tree, produced by an extended Press–Schechter formalism. In this model, the interstellar medium, stellar structures, and most importantly, BH seeds, co-evolve with each other through gravitational and dynamical interactions. BH seeds grow via accretions and mergers. In the merger scenario, after two galaxies have merged, the central BHs are dragged towards the new centre through dynamic frictions and form a binary. However, it is still unclear what mechanism drives the binary to GW- dominated separation. This question is known as the ‘last parsec problem’ \citep{begelman1980massive}. What is clear is that the time scale of this process is non-negligible. Therefore, the time delay between BH and galaxy mergers must include this part \citep[see][and references there in for details]{2016PhRvD..93b4003K}. 

Three ABH population models have been considered:

Model popIII: This model assumes light BH seeds are remnants of popIII stars, while accounting for the delays between BH and galaxy mergers. The inclusion of delays makes the model more realistic and more conservative. However, the typical eLISA event rates change by less than a factor of 2 when the delays are ignored. Therefore, the no-delay model of popIII is not considered here. The seed formation starts at $z$ = 20 and stops at $z$ = 15, and light seeds are formed with masses of $\sim 10^2 M_{\odot}$

Model D: This assumes heavy BH seeds are formed from the collapse of protogalactic disks with masses of $\sim 10^5 M_{\odot}$ at $z \sim 15-20$, while accounting for the delays between BH and galaxy mergers. The inclusion of delays makes the model more realistic and more conservative.

Model NOD: Same as model D, but without delay. This model should be considered as optimistic scenario.

\subsection{Signal-to-Noise}
\begin{figure}
    \includegraphics[width=\columnwidth]{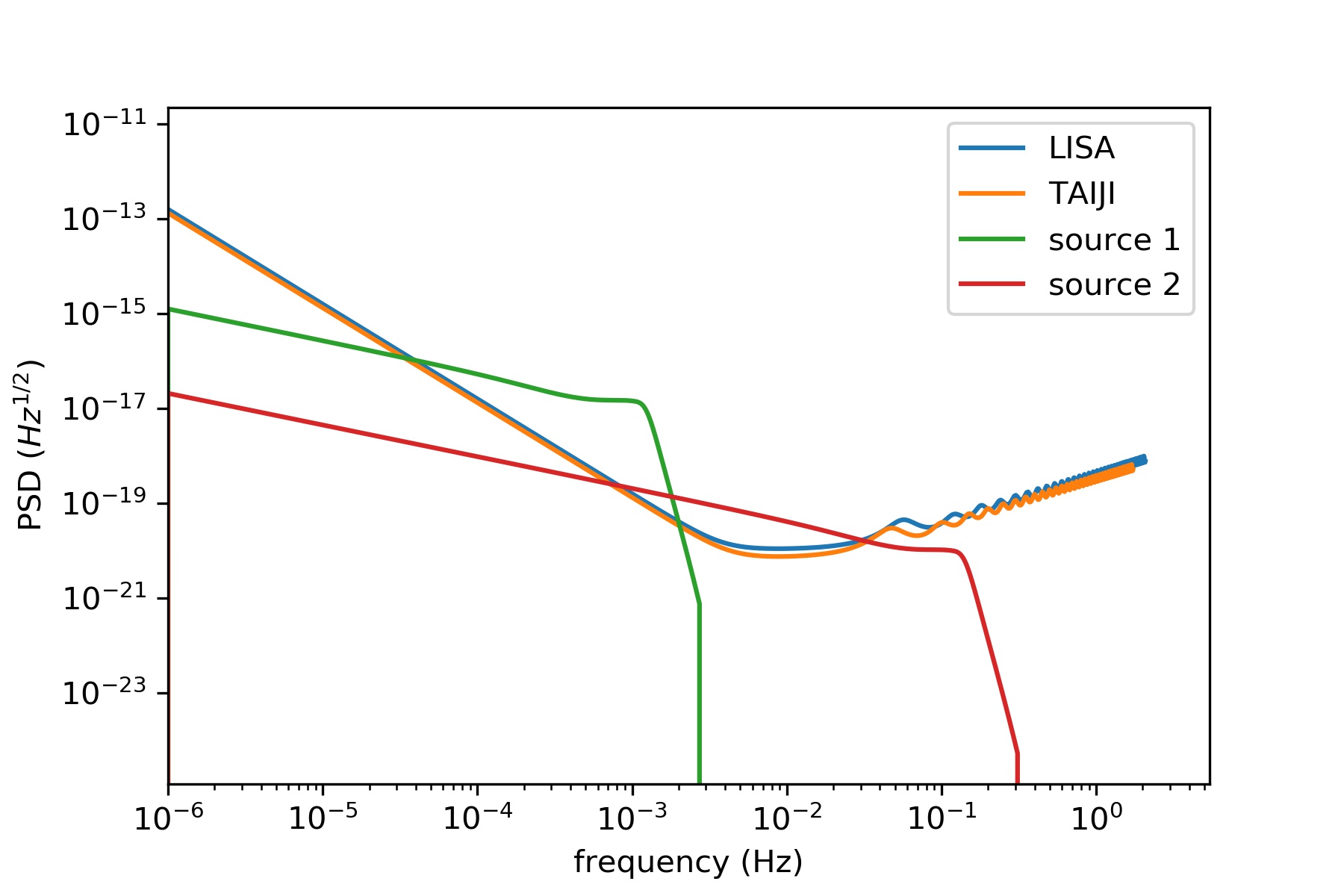}
    \caption{GW signals of two PBH binaries in frequency domain compared with LISA and Taiji (time-, sky- and polarisation-averaged) sensitivity curve.}
    \label{signals}
\end{figure}

The LISA interferometer is formed by three spacecraft with an arm-length 2.5 million km, interferometry sensing noise 10 pm/$\sqrt{\rm Hz}$ and test-mass force noice 6 fN/$\sqrt{\rm Hz}$;  \cite{2017arXiv170200786A}. As another space GW detector, Taiji, is proposed to be a LISA-like mission with a 3-million km arm length, sensing noise 8 pm/$\sqrt{\rm Hz}$ and the same test-mass force noice \cite{hu2017taiji}. In Fig.\ref{signals}, the time-, sky- and polarisation-averaged LISA and Taiji sensitivities are shown.

In this work, we adopt the official orbital configuration of Taiji, which leads the Earth by $\sim 20^\circ$, with a $+60^\circ$ inclined orientation (also called Taijip in  \cite{2021PhRvD.104b4012W,wanghan2021PRDb}). Given the parameters of each simulated BH binary, we calculate the response and SNR of LISA and Taiji, respectively. In the following analysis, we do not use the time, sky and polarization sensitivities. We use the power spectral density (PSD) of LISA and Taiji and calculate SNR etc. considering polarization angles and sky location line of sight by following \citet{2020PhRvD.102b4089W,wang2021observing,wanghan2021PRDb,2021PhRvD.104b4012W} (for a comprehensive description, see the content around Eq. (11) in \cite{2021PhRvD.104b4012W}) and top two panels in Fig. 2 inside for the instantaneous sensitivities on the sky map for the LISA mission and LISA-Taiji networks. The SNR $\rho_{\rm joint}$ of the joint network is calculated as:
\begin{equation}
    \rho_{\rm joint}=\sqrt{\rho_{\rm LISA}^2 + \rho_{\rm Taiji}^2}.
	\label{SNR}
\end{equation}

where $\rho_{\rm LISA}$ and $\rho_{\rm Taiji}$ are the SNR of LISA and Taiji mission respectively.

\section{Primordial Black Holes}\label{sec.PBH}
According to \citet{2021JCAP...05..003D}, LISA and other future space- borne detectors could detect up to hundreds of BH mergers at redshift $z$ > 30, which can therefore be confidently identified as being primordial. Therefore, we focus on the improvement of PBH detectability, especially on mass and distance/redshift.
\subsection{PBH population}
In this part, we generate a population of PBH binary and calculate their coalescence time. We adopt the PBH formation mechanism and initial constraint on parameters from \citet{2016PhRvL.117f1101S}. We consider a uniform initial mass function over a broad mass range of $10^2-10^9  M_{\odot}$, i. e., the probabilities of different initial masses are equal. For simplicity, we only consider equal-mass binaries. According to \citet{2016PhRvL.117f1101S}, a pair of PBHs separated by a distance of x can only decouple from the expansion of the Universe and become gravitationally bound when
\begin{equation}
    x < f^{1/3}\overline{x},
	\label{boundary}
\end{equation}
where $f$ is the fraction of PBHs in dark matter, i.e. $\Omega_{\rm PBH}=f \Omega_{\rm DM}$, and the physical mean separation $\overline{x}$ of PBHs at matter-radiation equality at the redshift $z=z_{\rm eq}$ is given by
\begin{equation}
    \overline{x}=(\frac{M_{\rm PBH}}{\rho_{\rm PBH}(z_{\rm eq}) })^{1/3}=\frac{1}{(1+z_{\rm eq})f^{1/3}}  (\frac{8\pi G}{3H_0^2}  \frac{M_{\rm PBH}}{\Omega_{\rm DM}} )^{1/3}.
	\label{mean separation}
\end{equation}
On the other hand, a pair of newly decoupled PBHs would directly collide rather than form a binary, if there is no presence of a third PBH close to the pair. The third PBH closest to the BH pair affects the infall motion of the PBHs in the pair by giving them the tidal force. As a result, the head-on collision does not happen and the PBHs in the pair form a binary, typically having a large eccentricity. If we define $y$ as the physical distance to the third PBH at $z=z_{\rm eq}$, the major axis and eccentricity of the binary can be described as
\begin{equation}
    a=\frac{1}{f}  \frac{x^4}{\overline{x}^3}, e=\sqrt{(1-(\frac{x}{y})^6}.
\label{a and e}
\end{equation} 
By definition, $y>x$ and $y<\overline{x}$ must be satisfied. Now, the coalescence time can be calculated following \citet{1963PhRv..131..435P} as

\begin{equation}
    t=Q a^4 (1-e^2)^{7/2},   
    Q=\frac{3}{170} (G M_{\rm PBH} )^{-3}.
\label{t}
\end{equation} 
\subsection{Gravitational wave simulation}
Now we focus on PBH binaries that coalescence at redshift $z$ = 20-50 and simulate GW signals from them. We use PYCBC \citep{alex_nitz_2021_5347736} to generate signals. The phenomenological waveform model we adopt is IMRPhenomPv2. To calculate SNR, we adopt the PSD data of LISA and Taiji from LISA official website\footnote{\url{http://lisa.nasa.gov/}} and \citet{hu2017taiji}, respectively, averaging over the sky position and polarization. Orbital-plane inclination is also averaged following the procedure of \citet{Robson_2019}

\subsection{Event rate and detection rate}
The event rate is calculated following \citet{2021arXiv210613769D}. For simplicity, we ignore the reduction of the PBH merger rate arising from interactions with surrounding inhomogeneities in the dark matter fluid and neighbouring PBHs around the binary formation epoch. Therefore, the event rate for equal-mass PBH binaries is
\begin{equation}
    R(M_{\rm tot})=\frac{5.7\times 10^6}{{\rm {\rm Gpc}}^3 {\rm yr}} f^{53/37} (\frac{t}{t_0})^{-34/37} (\frac{M_{\rm tot}}{M_{\odot}})^{-32/37}  ,
\label{R}
\end{equation} 

where $t_0$ is the current age of the universe. 
Now we can calculate the detection rate following  \citet{2021arXiv210613769D}. The detection rate is given as
\begin{equation}
    N_{\rm det}=\int dz dM_{\rm tot} \frac{1}{1+z}  \frac{dV_c(z)}{dz}  R(M_{\rm tot}) p_{\rm det} (M_{\rm tot},z)
\label{N}
\end{equation}

where $\frac{dV_c(z)}{dz}$ is the comoving volume per unit redshift, and the factor $p_{\rm det} (M_{\rm tot},z)$ accounts for the probability of detection of a binary, averaged over the source orientation, as a function of the signal-to-noise ratio (SNR). The exact form is given in the appendix of \citet{2021JCAP...05..003D}. For simplicity, we average over sky position, orbital-plane inclination, and polarization angle, and the probability is therefore simply a Heaviside step function $\Theta$
\begin{equation}
    p_{\rm det} (M_{\rm tot},z)=\Theta[\rho(M_{\rm tot},z)-\rho_{\rm thr}] \,,
\label{p}
\end{equation}
where we set the SNR threshold $\rho_{\rm thr}=10$, and  $\rho(M_{\rm tot},z)$ is given in the last section.

\subsection{Parameter estimation}
To compare the ability of PBH parameter estimation capabilities of the LISA-Taiji joint network with LISA alone, we use the Fisher information matrix to calculate errors of each parameter estimated by LISA and the joint network.

The detail of the Fisher matrix method can be found in \citet{di2019black}, and similar materials. The general idea of this method is to approximate the distribution of parameters around their maximum likelihood values as a multivariate Gaussian distribution of the form
\begin{equation}
    p(\boldsymbol{\theta}| \boldsymbol{d}) \propto e^{\delta \theta^{i} \Gamma_{ij} \delta \theta^{j}/2}
\label{ptheta}
\end{equation}
when data set $\boldsymbol{d}$ is given. Here $\delta\theta^{i}$ is the deviation of the $i$-th parameter $\theta^{i}$ from its maximum likelihood value, and $\Gamma_{ij}$ is the Fisher information matrix, whose inverse serves as the covariance matrix. 

The Fisher information matrix can be generally written as
\begin{equation}
    \Gamma_{ij} = (\frac{\partial{\tilde{h}}}{\partial{\theta^i}} | \frac{\partial{\tilde{h}}}{\partial{\theta^j}}),
\label{Gamma}
\end{equation}
where
\begin{equation}
    (g | h) = 4{\rm Re}\int_{0}^{\infty} \frac{g^*(f)h(f)}{S_{n}(f)} df,
\label{int}
\end{equation}
and $S_{n}(f)$ is the noise power spectrum. If the frequency-domain GW waveform $\tilde{h}$ can be written as $\tilde{h} = A e^{i\Phi}$, where A is the amplitude and $\Phi$ is the phase, then the exact form of Fisher matrix is
\begin{equation}
    \Gamma_{ij} = 4\int_{0}^{\infty} 
    \frac{
          \frac{\partial{A}}{\partial{\theta^i}} \frac{\partial{A}}{\partial{\theta^j}}
          +A^2\frac{\partial{\Phi}}{\partial{\theta^i}} \frac{\partial{\Phi}}{\partial{\theta^j}}
          }
    {S_{n}(f)} df.
\label{FIM}
\end{equation}
The Fisher matrix of the joint network is defined as the sum of the Fisher matrix of LISA and Taiji.

In this work, we consider 6 parameters: total mass $M_{\rm tot}$ (of two equal-mass BHs), luminosity distance $D_L$, sky location(longitude $\phi$ and latitude $\theta$), polarization angle $\Psi$, and inclination angle $\iota$. only consider equal-mass systems so mass ratios are fixed at 1. For simplicity, the spin of each BH is set to 0, as well as the coalescence time and phase. The true values of total mass and luminosity distance are randomly chosen from the LISA ‘waterfall’ with redshift $z > 20$, and other parameters are randomly chosen over all possible values. To our purpose, we focus on total mass and luminosity distance only in the result. The true values are shown in Table \ref{true_value}.

\begin{table}
\centering
    \resizebox{\columnwidth}{10mm}{
    \begin{tabular}{lccccccc}\hline
        &$\phi$(rad)&$\theta$(rad)&$\Psi$(rad)&$\iota$(rad)&$D_L$(Mpc)&$M_{\rm tot}(M_{\odot})$&z  \\\hline
        source1&4.27&1.04&3.14&0.00&299561.3&563801.4&25.4 \\
        source2&1.97&1.45&3.03&1.89&382408.5&3990.52&31.78 \\

    \end{tabular}
    }
    \caption{True values of parameters of 2 imaginary PBH binaries chosen for parameter estimation and their redshift.}
    \label{true_value}
\end{table}

\section{Results}\label{sec.result}
\subsection{Astrophysical Black Holes}

The sensitivities of LISA and Taiji are similar, when they receive
GW signals from the same binary, and if SNRs of each detectors are
similar, therefore according to equation (\ref{SNR}), one may think that the joint network would improve the SNR of LISA/Taiji by a factor of $\sqrt{2}$.

\begin{figure}
    \begin{subfigure}{\columnwidth}
        \includegraphics[width=\columnwidth]{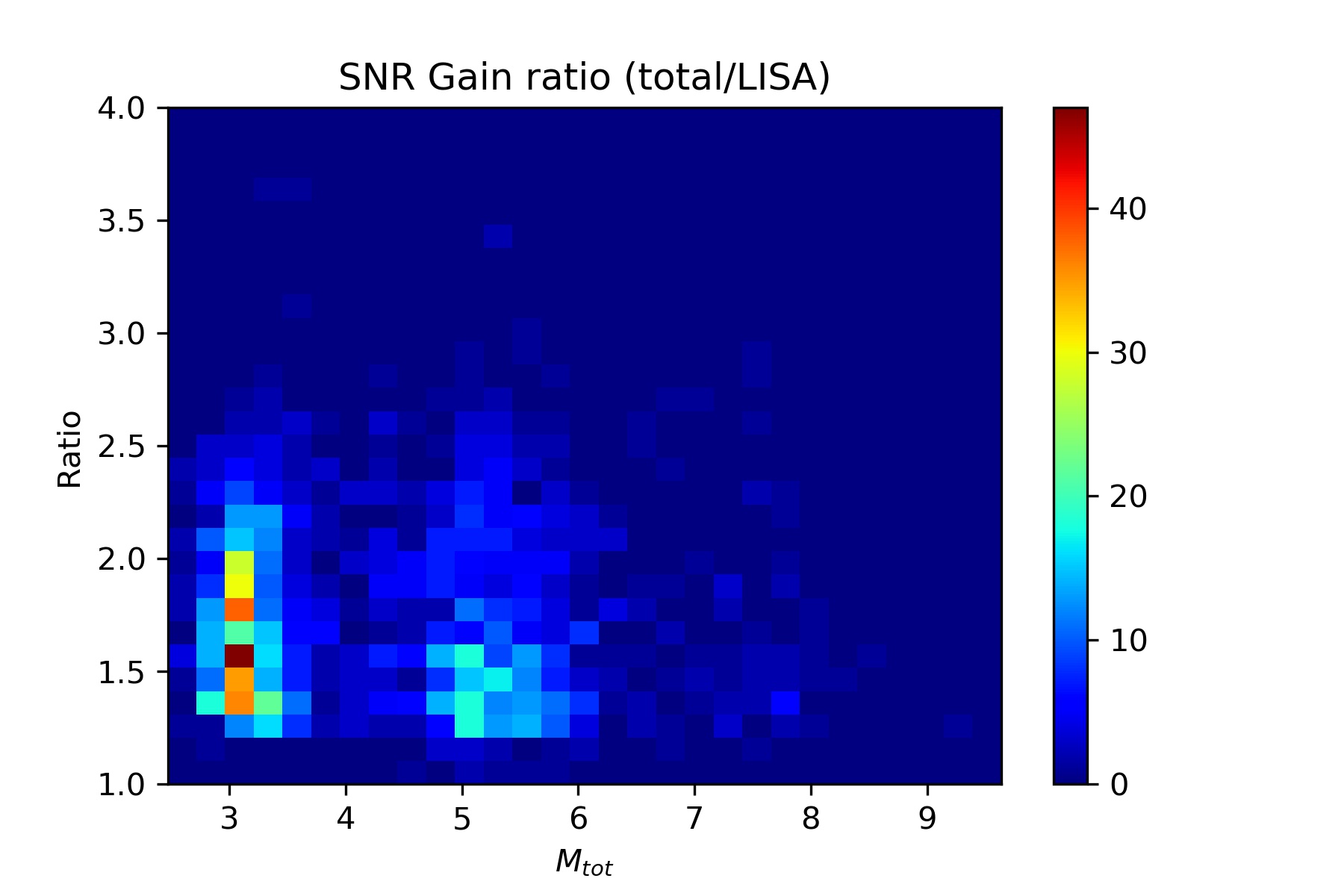}
    \end{subfigure}
	\begin{subfigure}{\columnwidth}
        \includegraphics[width=\columnwidth]{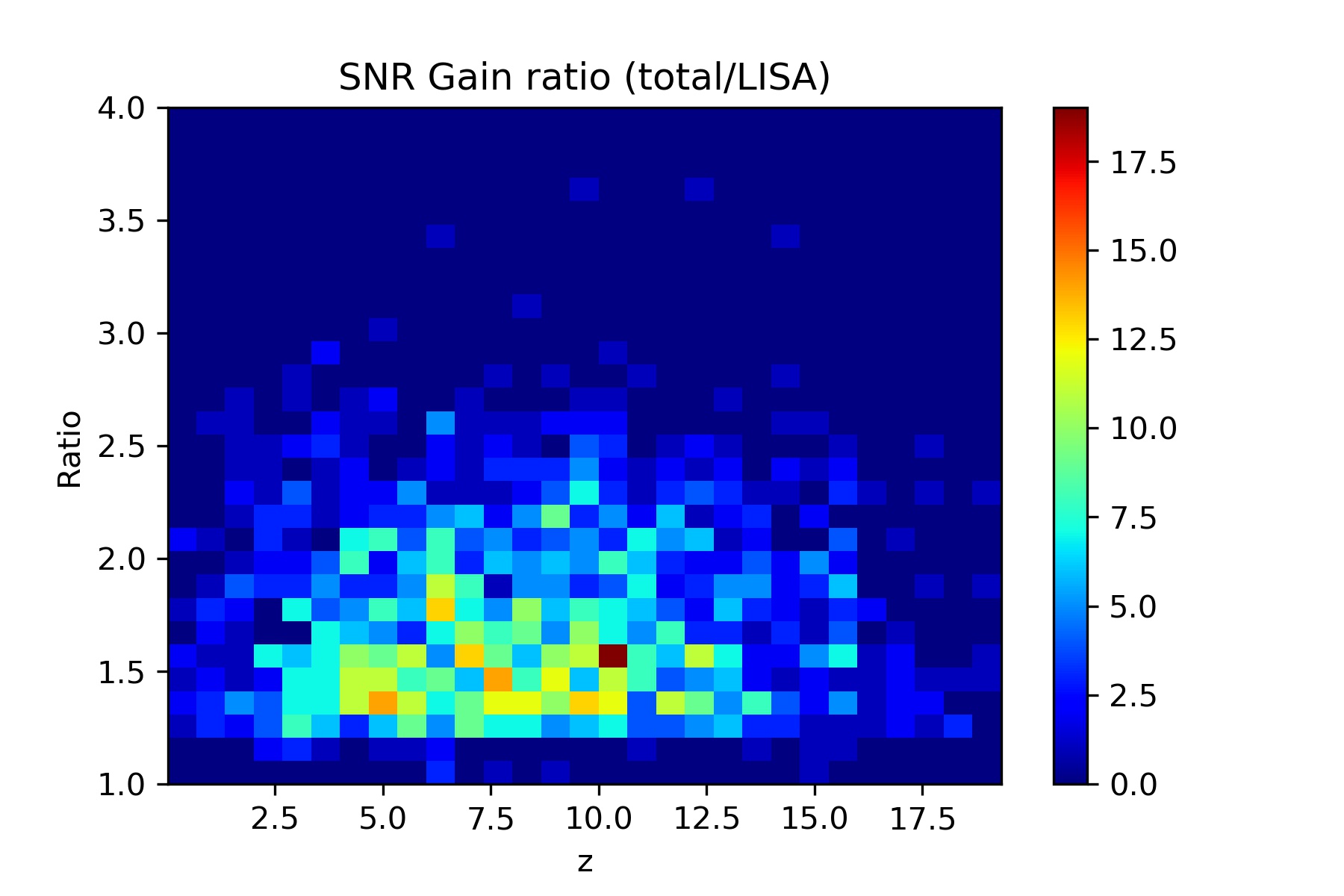}
    \end{subfigure}
    \caption{The ratio of SNR ($\rho_{\rm joint}/\rho_{\rm LISA}$) against total mass (top) and redshift (bottom) for all simulated BH binaries. The color bar indicates the number of binaries. Here, all three models of BH seeds are included.}
    \label{SNR ratio}
\end{figure}

\begin{figure}
    \begin{subfigure}{\columnwidth}
        \includegraphics[width=\columnwidth]{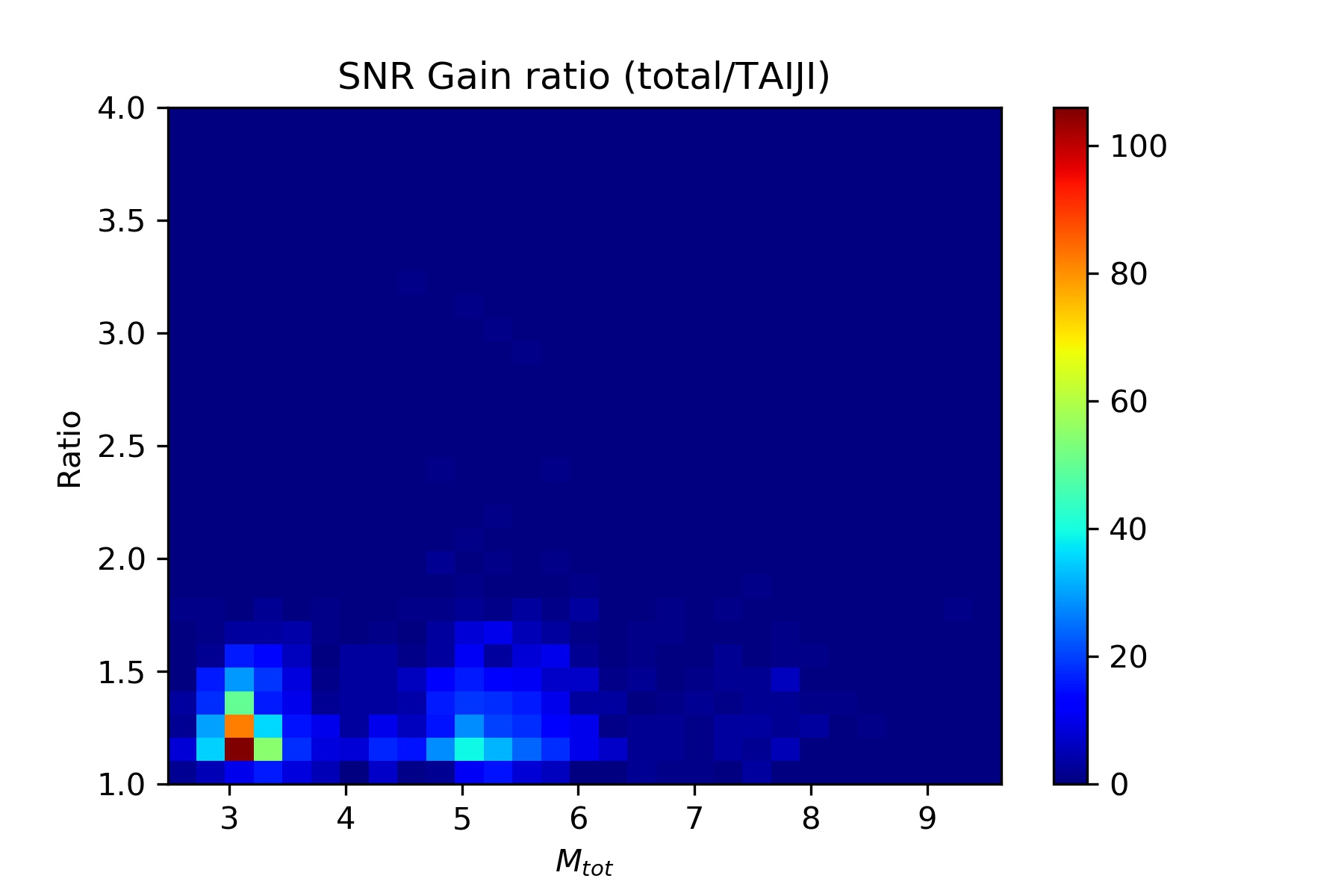}
    \end{subfigure}
	\begin{subfigure}{\columnwidth}
        \includegraphics[width=\columnwidth]{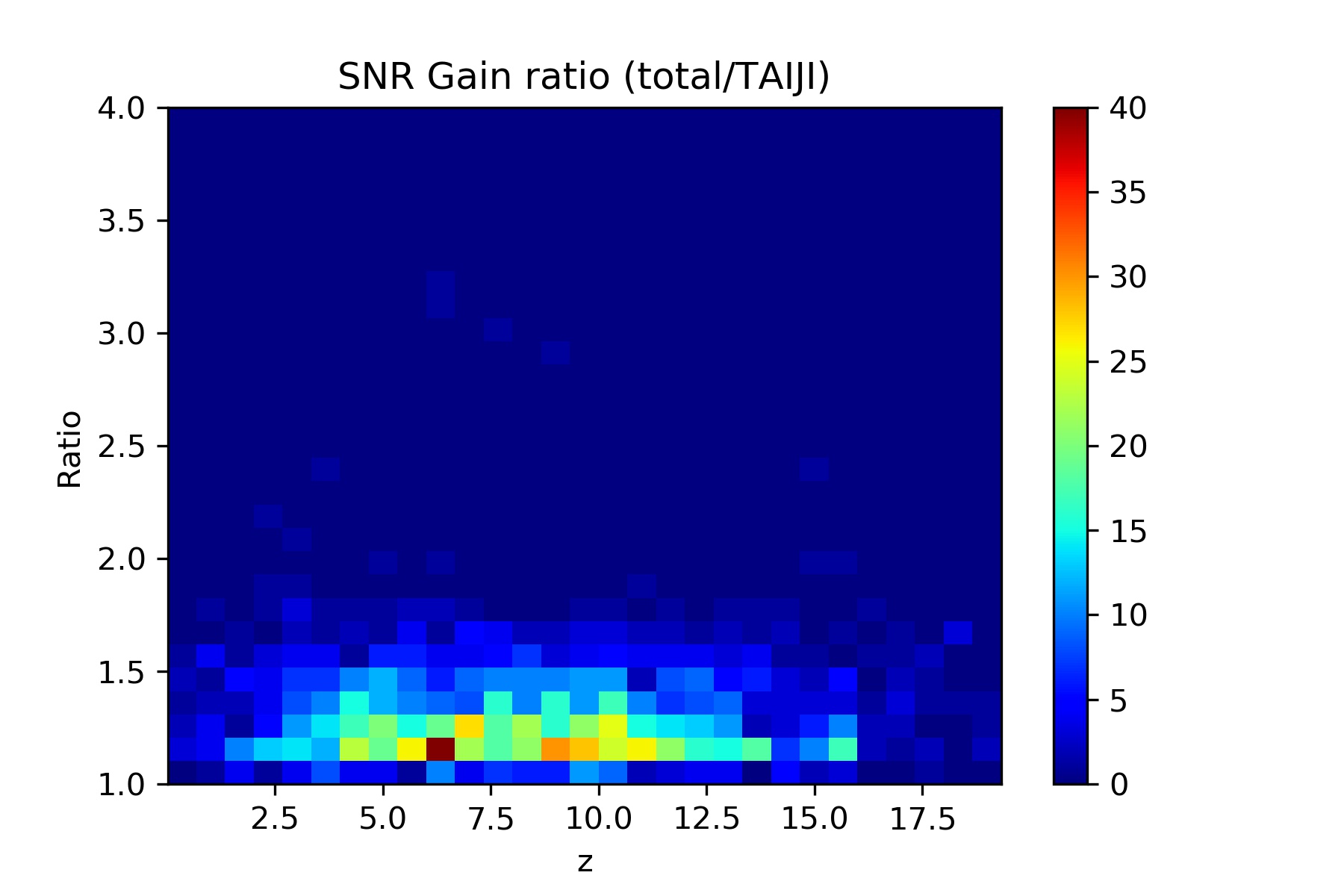}
    \end{subfigure}
    \caption{As the same as Fig. \ref{SNR ratio}, but the ratio of SNR ($\rho_{\rm joint}/\rho_{\rm Taiji}$) against total mass (top) and redshift (bottom) for all simulated BH binaries.}
    \label{SNR ratio_T}
\end{figure}

However, Fig.\ref{SNR ratio} shows the ratio of SNR of joint network and LISA (i.e., $\rho_{\rm joint}/\rho_{\rm LISA}$) over the binary’s total mass and redshift. When compared with LISA alone, the joint network would mostly improve the SNR of a single binary by a factor of $\sim 1.5$. The best cases show an improvement by factors about 3. The reason is that,
on the one hand, the sensitivity of Taiji is slightly better than LISA because of lower optical path noise requirement and longer arm-length, so the improvement on LISA should be slightly higher than $\sqrt{2}$ in average. On the other hand, two missions are not exactly the
same and are sensitive to different parameter spaces. Therefore, for a given binary in parameter space, two missions with high and low sensitivities together could result in a better improvement than $\sqrt{2}$ on the lower one. The joint network can enlarge the sensitivity range of each independent mission. The alternative Taiji orbital configuration induce that the interferometers have misaligned antenna pattern along their orbit compared to LISA, so for a given source in a given position in the sky Taiji (LISA) might detect the source with a higher SNR. Our analysis contained differences in the antenna pattern due to the different orbit travelled by the interferometers ( \cite{2020PhRvD.102b4089W,wang2021observing,wanghan2021PRDb,2021PhRvD.104b4012W}). In this paper, for the aim of revealing the network’s advantages, we focus on the improvement of network comparing with LISA throughout the work. As a comparison, we also plot the ratio of SNRs between network and Tajji (\ref{SNR ratio_T}), one can see the improvement is not as good as LISA.

\begin{figure}
    \includegraphics[width=\columnwidth]{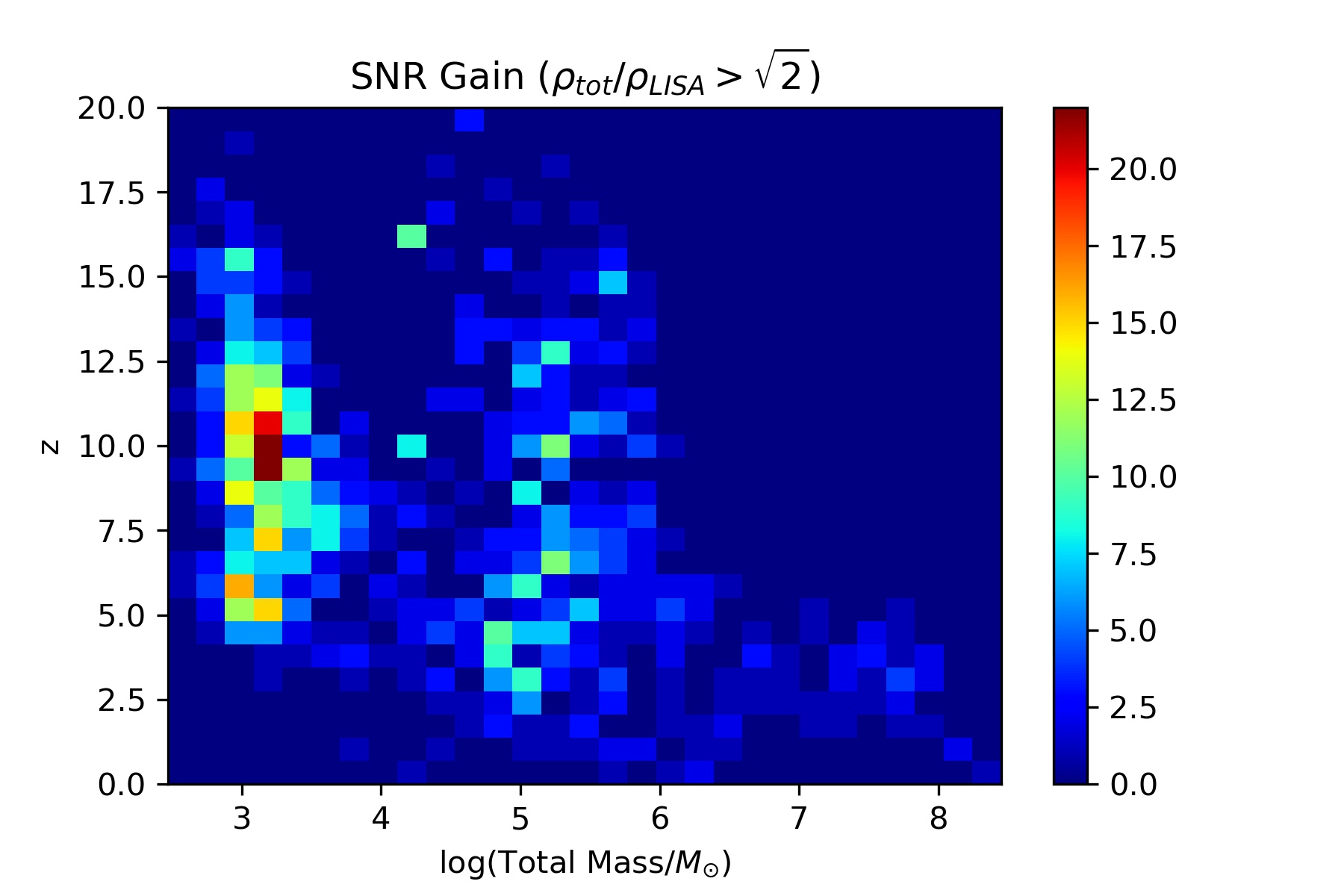}
    \caption{Binary counts on total mass and redshift. Only sources with SNR improvement greater than $\sqrt{2}$ on LISA are counted.}
    \label{overall}
\end{figure}

Employing the model developed by \cite{barausse2012evolution,2016PhRvD..93b4003K}, we generate the catalog for the binary BHs. We collect all the GW sources in the LISA frequency band ($10^{-5}-1$ Hz) in a 5 yr observation. The top plots in Figs. (\ref{SNR ratio}) show two gathering regions on the total mass. They correspond to light seeds (with mass $\sim 10^3 M_{\odot}$) and heavy seeds (both delayed and non-delayed, with mass $\sim 10^5 - 10^6 M_{\odot}$). For a better view, Fig. \ref{overall} shows an overall view in total mass-redshift dimension for binaries with SNR improvement on LISA better than $\sqrt{2}$. Indeed, the improvement is prominent on both light seeds and heavy seeds. Now we discuss two populations separately.

\subsubsection{SNR improvement for light seeds}
As \citet{2021MNRAS.500.4095V} pointed out, in the light seed scenario, there exists a group of "growing light seeds" with total mass $\sim 10^3 - 10^4 M_{\odot}$ and redshift $10 < z < 20$, which lies on the edge of LISA horizon. These merging light seed binaries, if detected, could shed light on the light seed formation and growth mechanism. Therefore, for light seeds, we focus on the ones with $z$ > 10.

\begin{figure}
    \includegraphics[width=\columnwidth]{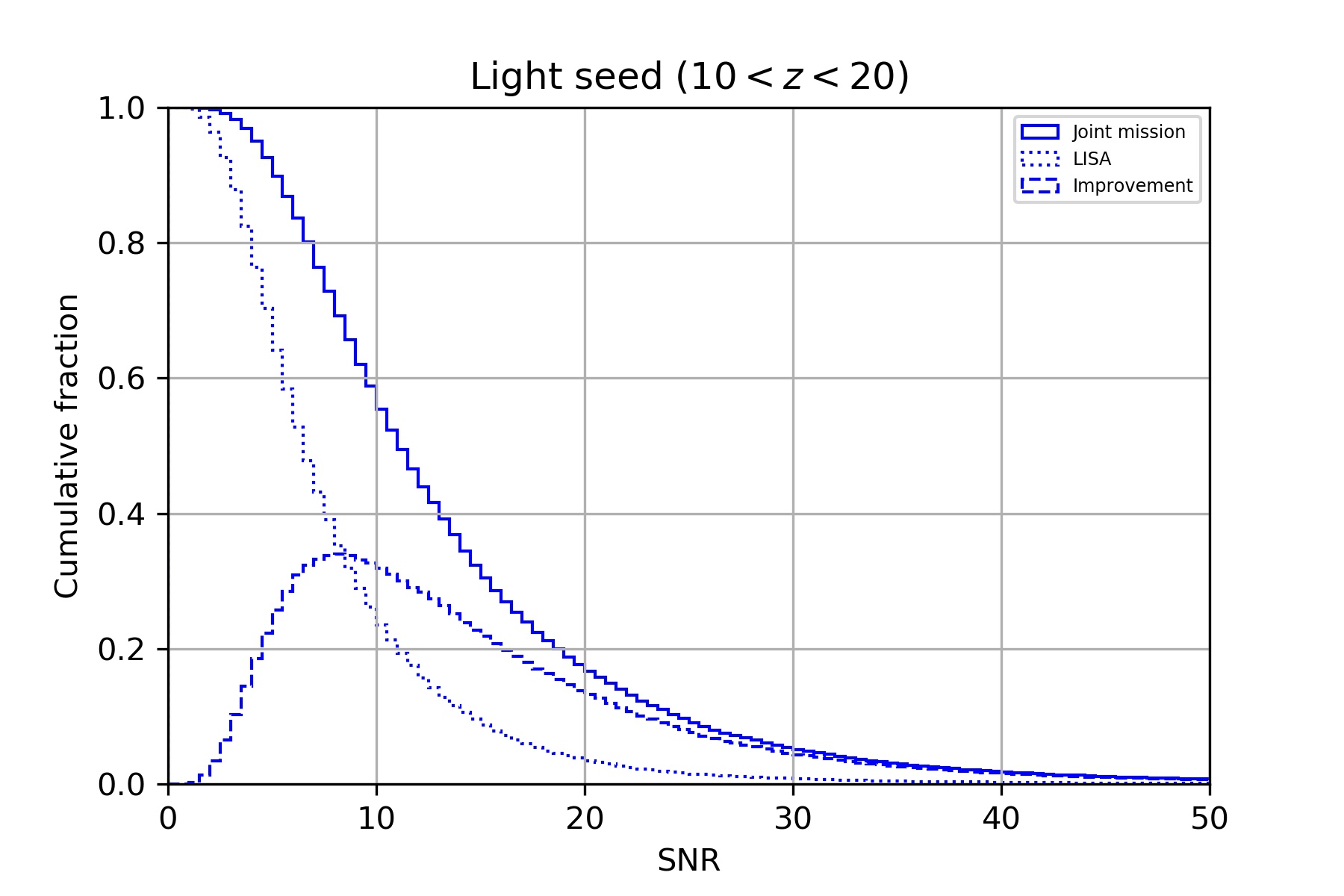}
    
    \caption{The inverse cumulative histogram of SNR of light seeds with $10 < z < 20$. The solid line indicates the SNR of the joint network, while the dashed line represents the SNR of LISA. The long dashed line stands for improvement, which is the subtraction of two histograms above, representing the number/fraction of light seed binaries that the joint network could add to LISA above certain SNR threshold.}
    \label{Light}
\end{figure}

Fig. \ref{Light} shows the reversed cumulative histogram of SNR of light seeds with $z$ > 10. The solid line represents the joint network, while the dashed line represents the LISA mission. The long dashed line represents how many more binaries/fractions the joint network can detect when compared with LISA above a certain SNR threshold. For example, the joint network can detect $\sim 55\%$ light seed binaries with SNR greater than 10, while LISA along can only detect $\sim 25\%$ above the same SNR. The improvement peaks at SNR 6-8, which is often considered as the threshold of determining whether a binary can be detected or not. The joint network could bring $\sim 35\%$ of light seeds, which is below the threshold in LISA, over the threshold and thus they can be detected. 

\begin{figure}
    \includegraphics[width=\columnwidth]{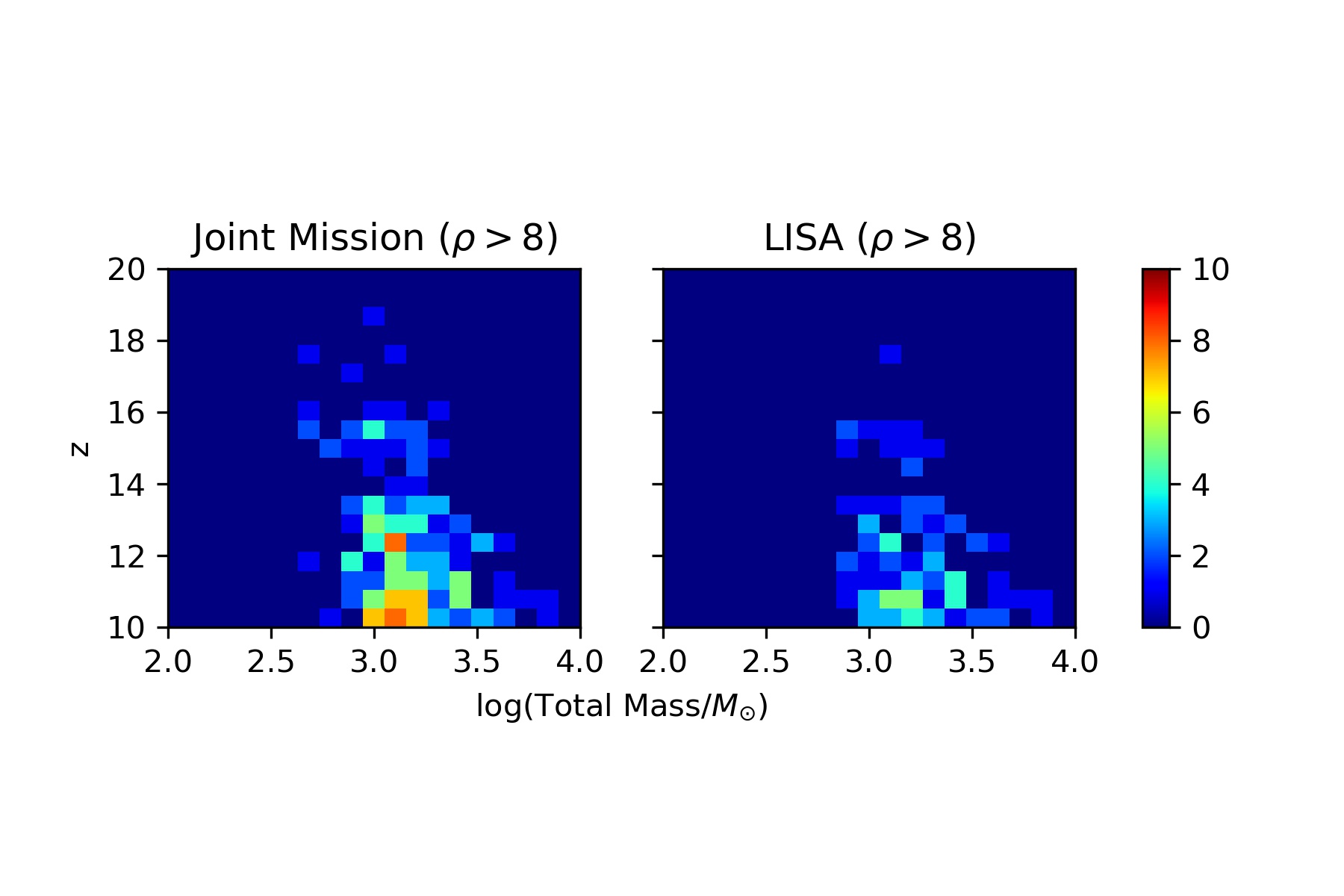}
    
    \caption{the redshift and total mass distribution of light seed binaries that can be detected by the joint network and LISA with SNR great than 8.}
    \label{areaA}
\end{figure}

It turns out that the joint network not only can improve the overall detectability of a single detector, but also detect light seeds in the parameter space where a single detector can hardly reach. Fig. \ref{areaA} shows the total mass and redshift distribution of detectable (i. e. $\rho$ > 8) light seed binaries. It is worth noticing that the joint network can detect binaries with $z$ > 17, which cannot be detected by LISA alone. This area lies on the edge of the LISA ‘waterfall’ (i.e. detection horizon of mass and redshift, see Fig. 3 of LISA proposal \citet{2017arXiv170200786A} for example). The joint network basically increases the size of the ‘waterfall’. Therefore, binaries on the edge can be detected. Light seeds with such high redshift are very close to their birth time, the cosmic dawn. These seeds quickly form merging binaries soon after the popIII stars collapsed. If detected, these binaries would be the ground-breaking evidence that light seeds form and grow in the high-$z$ and metal-poor environment, and the joint network would be crucial to the detection.

\subsubsection{SNR improvement for heavy seeds}
\begin{figure}
    \includegraphics[width=\columnwidth]{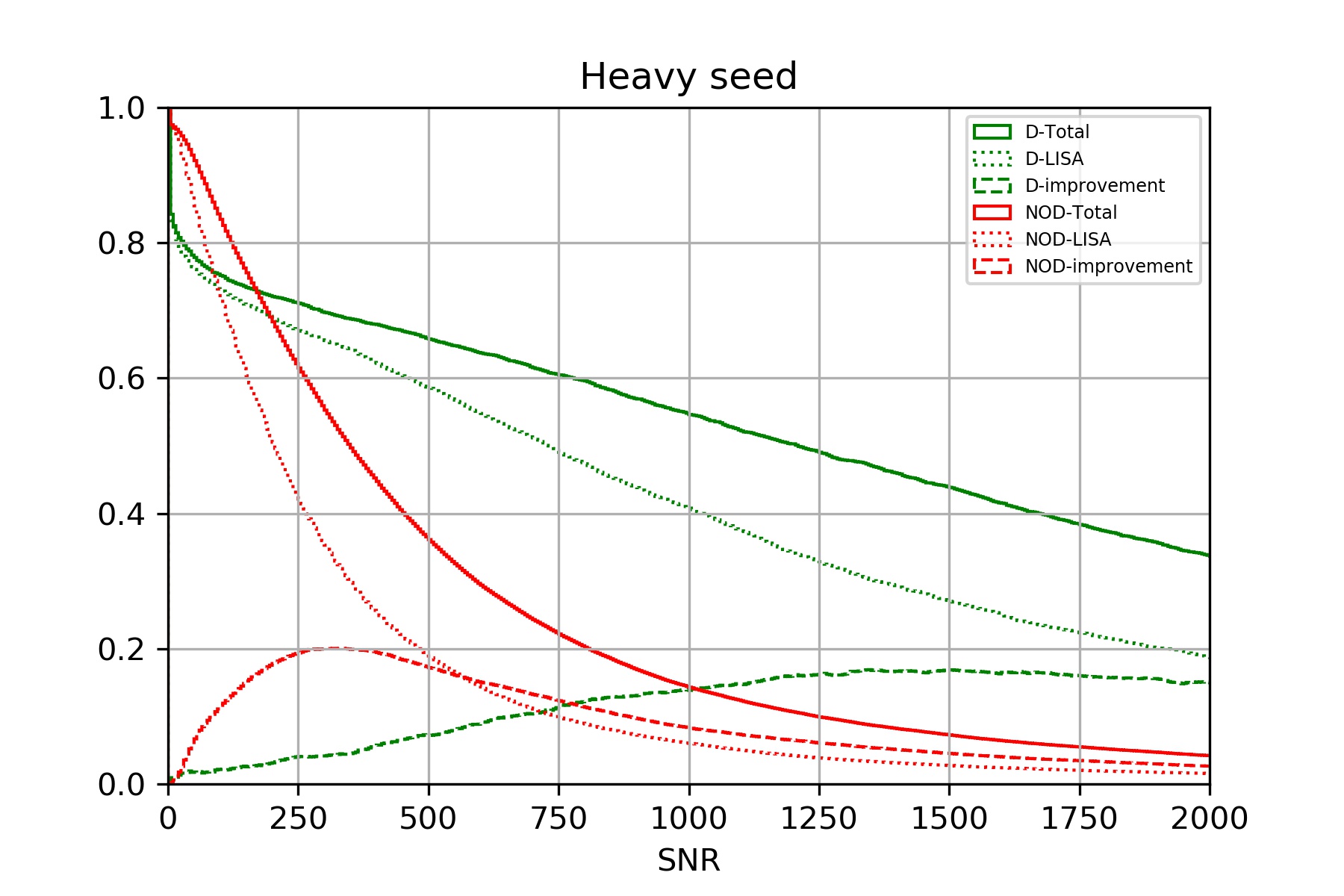}
    
    \caption{The inverse cumulative histogram of SNR of heavy seeds. The solid line indicates the SNR of the joint network, while the dashed line represents the SNR of LISA. The long dashed line stands for improvement.}
    \label{Heavy_seed}
\end{figure}

In the heavy seed scenario, BH mergers with delays happen at later times when compared with light seeds. Therefore, we pay attention to the whole redshift range (0–20). Fig.\ref{Heavy_seed} shows the inverse cumulative histogram of SNR of each heavy seed model. For the no-delay model, the improvement peaks at $\rho \sim 300$, indicating an improvement of 20\%. For the delayed model, the inverse cumulative histogram falls slower than the no-delay model, and the improvement peaks at $\rho \sim 1500$. The reason is that the merger of BHs happens at a later time in the history of universe, i.e., lower redshift. Therefore, the average SNR of this model are expected to be higher than the no-delayed model. In this work, we do not compare these two models. Instead, we treat them as the optimistic and conservative results of heavy seeds scenario, respectively. In this sense, both the results show a maximum improvement of $\sim 20\%$.

\begin{figure}
    \includegraphics[width=\columnwidth]{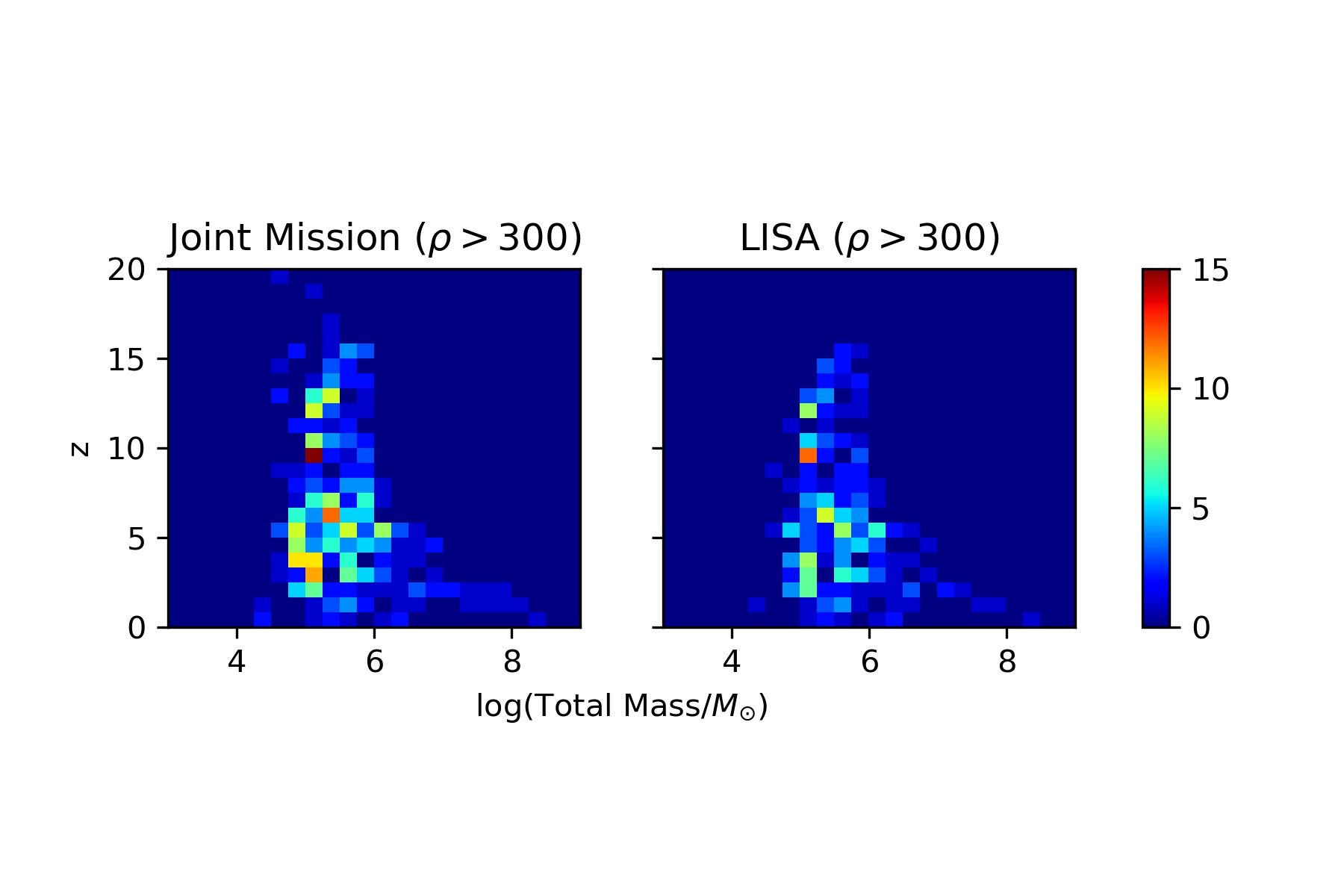}
    
    \caption{the redshift and total mass distribution of heavy seed binaries that can be detected by the joint network and LISA with SNR great than 300.}
    \label{areaB_300}
\end{figure}

\begin{figure}
    \includegraphics[width=\columnwidth]{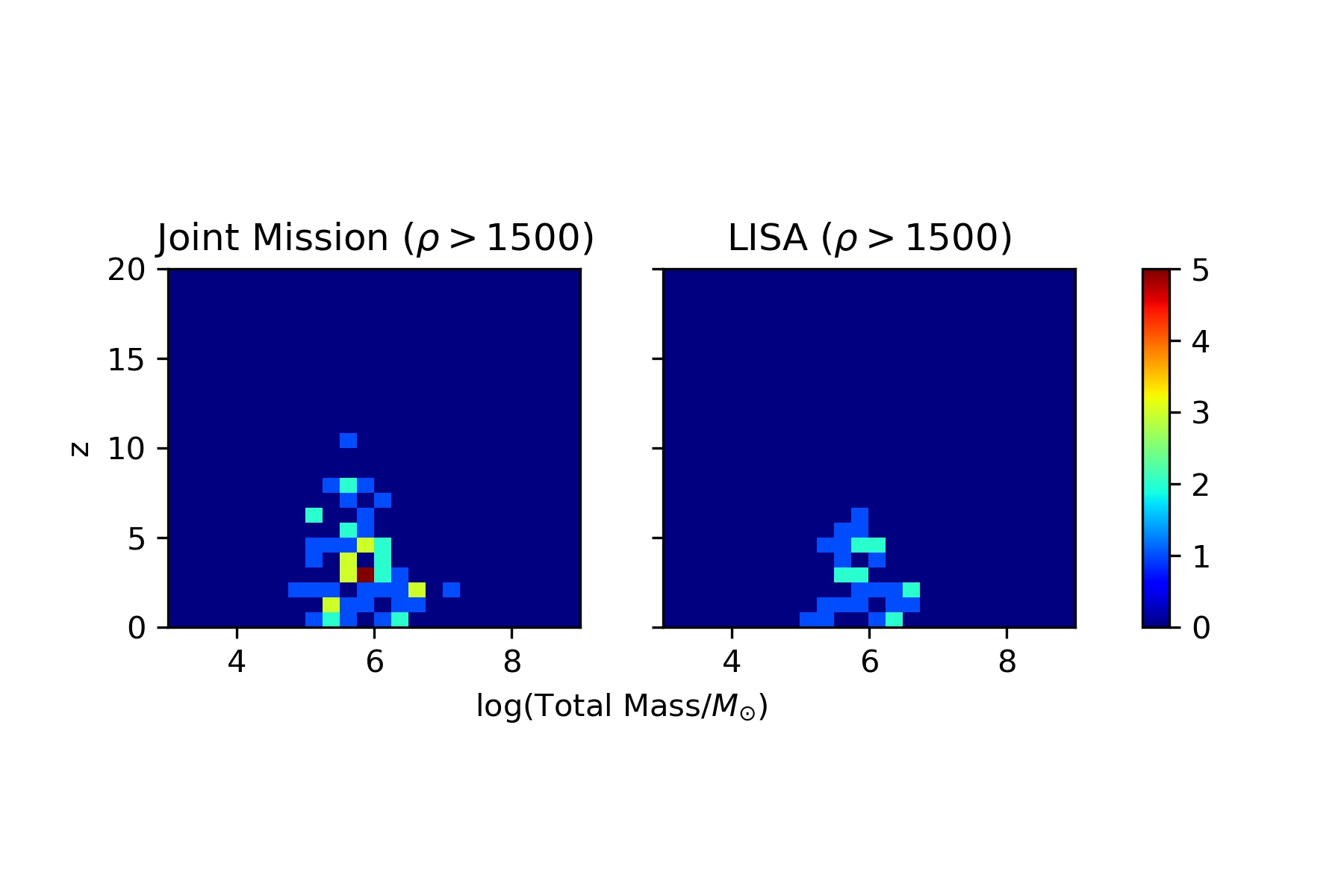}
    
    \caption{the redshift and total mass distribution of heavy seed binaries that can be detected by the joint network and LISA with SNR great than 1500.}
    \label{areaB_1500}
\end{figure}

The major difference between delayed and no-delay heavy seed binaries is the redshift at merger, which leads to different SNRs. Fig. \ref{areaB_300} and Fig. \ref{areaB_1500} shows the total mass and redshift distribution of binaries with SNR higher than 300 (1500), which is the improvement peak of the no-delay(delayed) model, in LISA and the joint network. At SNR threshold 300 (Fig. \ref{areaB_300}), most of the improvement that the joint network could bring are on no-delay binaries, which locate at $z > 10$. On the other hand, At SNR threshold 1500 (Fig. \ref{areaB_1500}), there are few no-delay binaries left in the joint network and LISA (<10\%). What are left are mostly delayed binaries with a lower redshift (z < 10).

In the heavy seed scenario, BHs already have masses $\sim10^5 M_{\odot}$ at their birth. The whole evolution of heavy seeds in our simulation lies within the horizon of LISA, so LISA alone can observe these binaries all. The joint network only improves SNR.

\subsection{Primordial Black Holes}

\begin{figure}
    \includegraphics[width=\columnwidth]{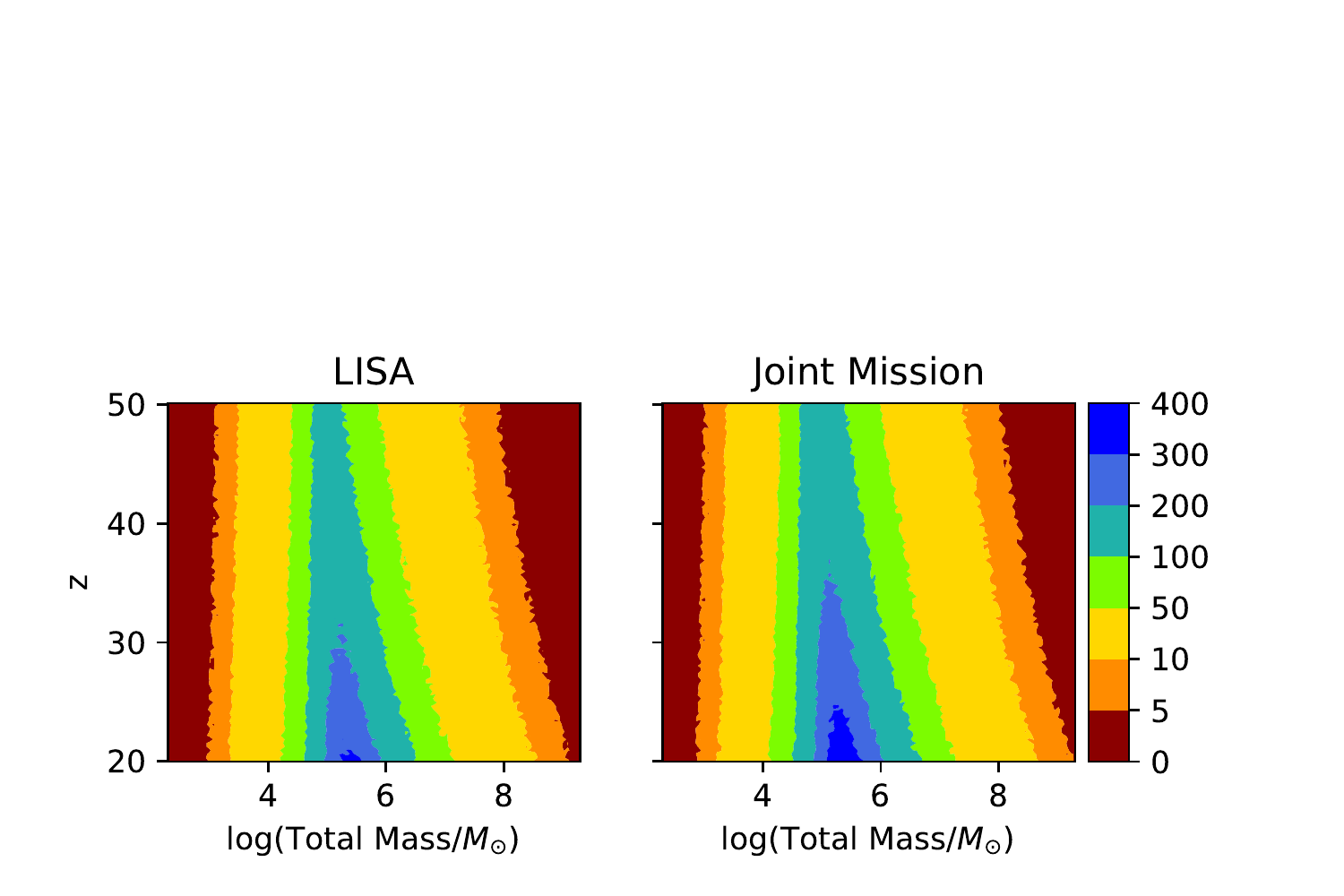}
    
    \caption{The distribution of SNR of PBH binaries over redshift and total mass. The color bar represents SNR level.} 
    \label{PBH}
\end{figure}

Fig. \ref{PBH} shows the SNR ‘waterfall’ (i.e. horizon over redshift and total mass) of LISA and joint network. Here, we calculate the SNR with equal-mass binaries because the mass ratio of PBH binaries is unfortunately unknown, and the waterfall in LISA L3 proposal by \cite{2017arXiv170200786A} adopted a constant mass ratio of 0.2. Note that according to our calculation of coalescence time, binaries with high eccentricity would coalescence first. Therefore, in this specific window, the initial eccentricity of PBH binaries is quite high (>0.99) at the moment of the binaries formed. This eccentricity remains even at merger since circularization might not be so efficient as the initial eccentricity is very large. Therefore, we employ SEOBNRE (a waveform model for an eccentric binary BH based on the effective-one-body-numerical-relativity formalism \citep{caohan2017PRD, yunhan2021PRD}) to generate the waveforms with random orbital eccentricity up to 0.6 before merger. For a fixed redshift, the smaller the total mass, the higher the event rate. The region of high event rate ($R > 1 {\rm Gpc}^{-3} {\rm yr}^{-1}$) intersect with the rising side of LISA and joint network ‘waterfall’, so one would expect high probability of detection. Especially, the joint network expands the LISA horizon towards the high event rate region, so it is more possible to detect such event with the joint network. 
\begin{figure}
    \includegraphics[width=\columnwidth]{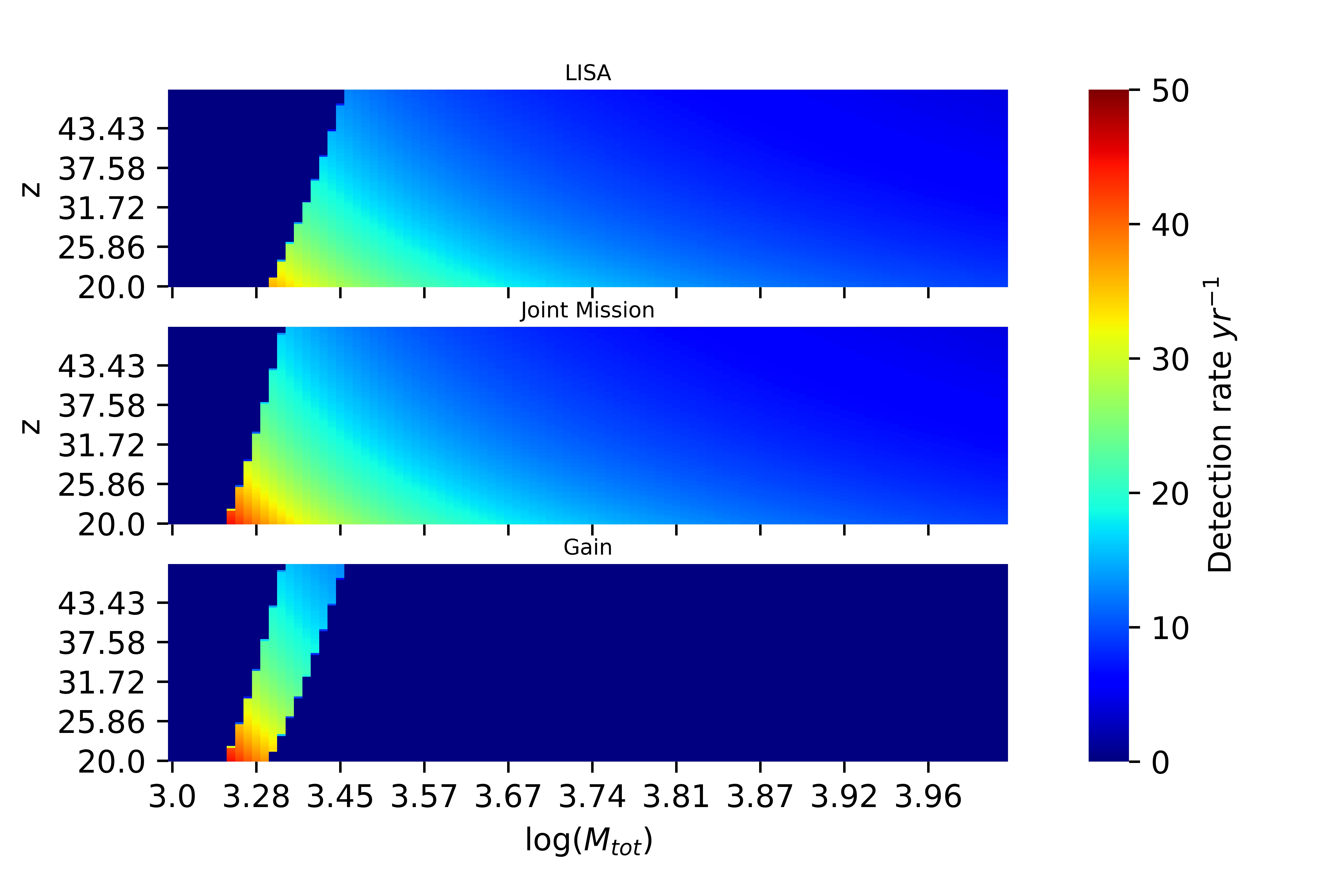}
    
    \caption{Detection rate of LISA (top), joint network (middle), and the improvement (bottom). The improvement is the difference in the detection rates of the joint network and LISA at each point.}
    \label{Detection rate}
\end{figure}

Fig. \ref{Detection rate} is the detection rate of LISA and the joint network. The bottom panel shows the improvement, which is the subtraction of the middle and top panels. The joint network can detect PBH binaries that LISA alone cannot observe, at a narrow band of total mass $10^{3.25}\sim 10^{3.45}$. Although it is narrow, the detection rate on this band is high, because of the high event rate in this area. Note that the high-mass end of the bottom panel is empty, because in equation (\ref{p}) we only consid- ered a step function, which functions only at the ‘waterfall’ boundary, and the detection rate is the same inside the ‘waterfall’ for both the joint network and LISA. The SNR is actually a function of not only in- trinsic parameters (mass and redshift) but also of extrinsic parameters (sky position, inclination, and polarization). However, we averaged over the extrinsic parameters in this work, which made SNR only a function of mass and redshift. Therefore, we note that this empty region may not be empty if extrinsic parameters are considered.
\begin{table}
    \centering
    \begin{tabular}{c|c|c|c}\hline
        
        &&$M_{\rm tot}(M_{\odot})$&$ln(D_{L}/Mpc)$  \\\hline
        \rule{0em}{10pt}
        LISA&source1&$563740.95^{+115364.87}_{-114775.63}$&$12.62^{+0.23}_{-0.30}$ \\
        \rule{0em}{10pt}
        &source2&$3990.57^{+24.97}_{-25.43}$&$12.85^{+1.83}_{-1.86}$ \\\hline
        \rule{0em}{10pt}
        joint network&source1&$563820.34^{+63210.24}_{-63247.12}$&$12.61^{+0.01}_{-0.01}$ \\
        \rule{0em}{10pt}
        &source2&$3990.52^{+14.07}_{-13.29}$&$12.86^{+0.12}_{-0.13}$ \\\hline

    \end{tabular}
    \caption{Total mass and luminosity distance estimations of two PBH binaries using Fisher matrix of LISA and the joint network respectively. Errors are calculated on $1-\sigma$ level.}
    \label{prameter_estimation}
\end{table}

\begin{figure}
    \includegraphics[width=\columnwidth]{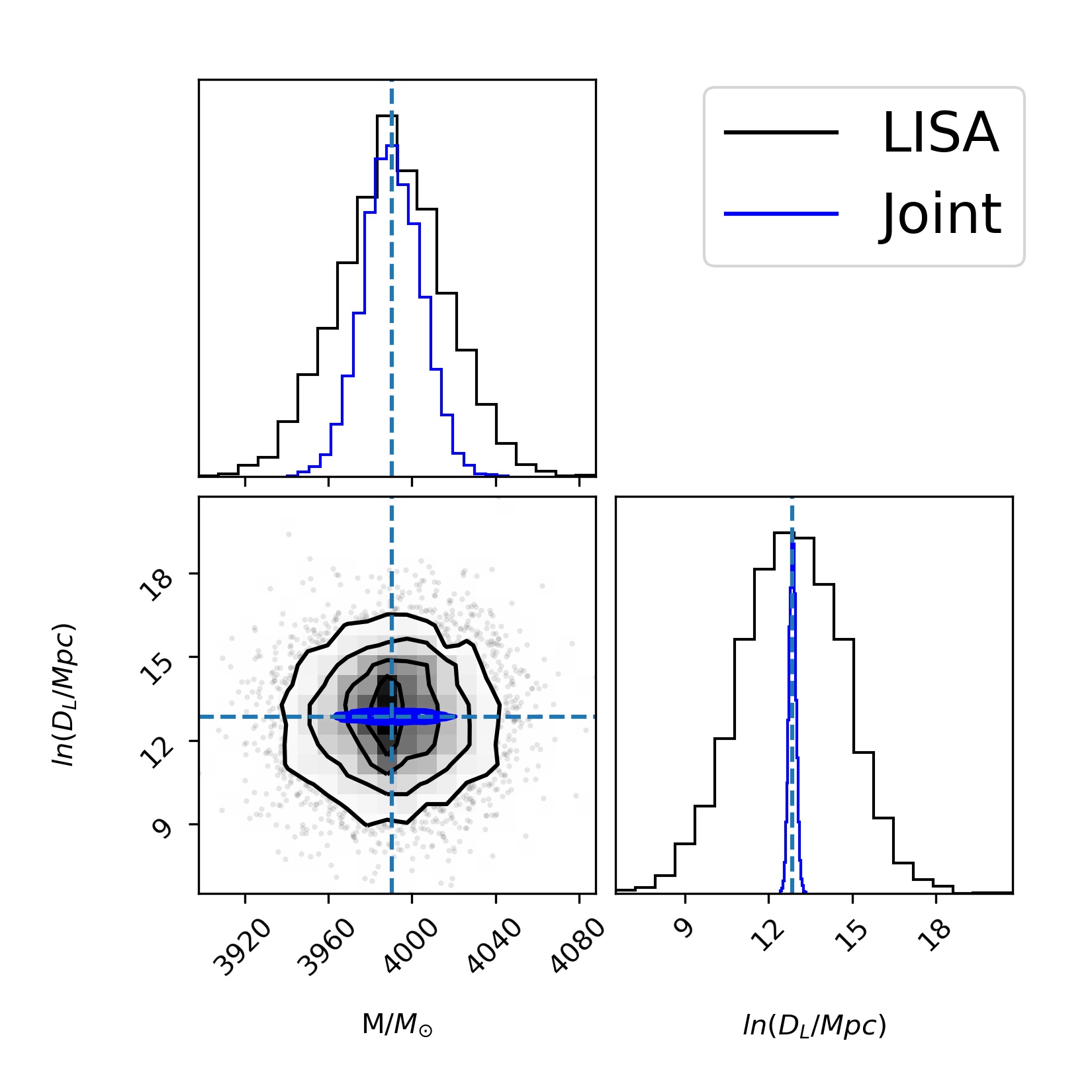}
    
    \caption{Luminosity distance and total mass estimation of LISA and joint network.}
    \label{corner}
\end{figure}

Fig. \ref{corner} shows the total mass and luminosity distance estimation derived from the Fisher matrix of LISA and the joint network, while the values and errors are shown in table \ref{prameter_estimation}. Here, we only show the plot of the second PBH binary for a better view. In Fig. \ref{corner}, there is a significant decrease in the uncertainty area covered by the joint network. Especially on luminosity distance, the uncertainty reduces by 93.2\%, suggesting strong ability of determining distance using the joint network. Most importantly, this binary with total mass of $\sim 10^3 M_{\odot}$ and $z \sim 30$ lies on the rising side of LISA ‘waterfall’, where the PBH merger event rate is high, and is more likely to be observed by the joined mission. Therefore, with the joint network, we have better chance of detecting PBH binaries merging, if they exist, and better determination of their distance and total mass, which would be a smoking gun in favour of PBH theory.

The LISA-Taiji network then improves the estimation of distance greatly with an uncertainty of only 1\%. On the opposite, the second binary enters LISA band at an early time of its inspiral, and has a smaller orbital period. Therefore, LISA can estimate the distance of the second binary with an uncertainty of 15\%. The LISA-Taiji network can improve the estimation to just 1\% uncertainty.

n addition, the Fisher matrix is only a crude way of estimation. It treats the posterior distribution as Gaussian. However, the approximation only holds when the value is very close to the maximum-likelihood value. With such high uncertainty, the Fisher matrix method clearly fails on estimation. A more precise way of estimating the distribution is through MCMC \citep[see][for example]{cornish2020black}. We leave this for future work.

\section{Conclusions}
In this work, we have discussed the joint network of future GW detectors LISA and Taiji. The joint network is expected to improve
the SNR of the individual detector by a factor of $\sqrt{2}$. However, Taiji is slightly better than LISA, so we find that the improvement on LISA is concentrated on 1.5, with the highest case being 3, over redshifts and masses. 

For ABHs, we have discussed two BH seeding mechanisms, and how better the joint network can observe them. For the light seed, the joint network enlarges the horizon of LISA, so binaries with higher distance and/or smaller total mass can be observed by the joint network. The growing light seeds, i.e. light seeds that are just formed, with total masses of $\sim 10^3 M_{\odot}$ and redshift $z \sim 15-20$, can be observed by the joint network, while LISA can hardly see. If they are detected, it would be groundbreaking, unveiling the light seed forming and dynamical pairing mechanisms. The heavy seeds, however, locate at the centre of LISA horizon. All the joint network observations can be observed by LISA alone. Yet, the joint network improves the SNR, which could provide better parameter estimations.

For PBHs, we focus on redshift $20 < z < 50$, where the ABHs are not yet formed, so if a BH binary is detected, it would most probably be of primordial origin. We calculate the event rate of PBH and detection rates of the joint network and LISA. The joint network increases the horizon towards the lower mass end by $10^{3.25} \sim 10^{3.45} M_{\odot}$. However, the event rate rises on the lower mass end. This narrow band of improvement is actually high detection rate region, so the joint network brings more chance of detecting PBH binaries at high redshift. We also estimate parameters of PBH binaries using Fisher matrix. The joint network decreases the uncertainty significantly, especially on luminosity distance. Therefore, with the joint network, if we receive a signal from that region, we can have more confidence in determining it as a high-z source, which is more likely to be PBH binaries. This would be a smoking gun in favour of PBH theory.

In general, the joint network enlarges the mass-redshift ‘waterfall’ of individual missions, and it is more interesting on the left rising side (i.e. lower mass end) both for ABH and PBH theories. The next-generation ground-based detectors ET and CE cover the lower mass end ($10 \sim 10^3 M_{\odot}$) \citep{Ng2021arXiv210807276N}, and intersect with the joint network. LISA and Taiji can in principle detect binary mergers with masses of $10^3\sim 10^6 M_\odot$ which can either form in the early universe from PBHs or ABHs. Therefore, it would be interesting to investigate the detection capability of a ‘Grand Union’ of all future GW detectors.
\section*{Acknowledgements}
This work is supported by The National Key R\&D Program
of China (Grant No. 2021YFC2203002), NSFC (National Natural Science Foundation of China) No. 11773059 and No. 12173071. W. H. is supported by CAS Project for Young Scientists in Basic Research YSBR-006. We thank Xingyu Zhong and Chen Zhang for discussion and support on the use of analysis tools used in this work. We thank Gang Wang for crucial help on data analysis.  We appreciate for the useful comments from the anonymous referees.
\section*{DATA AVAILABILITY}
The data underlying this article will be shared on reasonable request to the corresponding author Wen-Biao Han.



\bibliographystyle{mnras}




\appendix


\bsp	
\label{lastpage}
\end{document}